\documentclass[runningheads]{llncs}
\usepackage[utf8]{inputenc}
\usepackage{graphicx}
\usepackage{caption}
\usepackage{subcaption}
\usepackage{comment}
\usepackage{amssymb,amsmath}
\usepackage{cite}
\usepackage[pdfpagelabels,colorlinks,allcolors=blue]{hyperref}

\usepackage{booktabs}
\usepackage{makecell}
\usepackage{tabularx}
\usepackage[ruled,vlined,noend]{algorithm2e}

\usepackage{todonotes}
\usepackage{xspace}
\usepackage{xcolor,colortbl}
\usepackage{float}

\renewcommand{\orcidID}[1]{\href{https://orcid.org/#1}{\includegraphics[scale=.03]{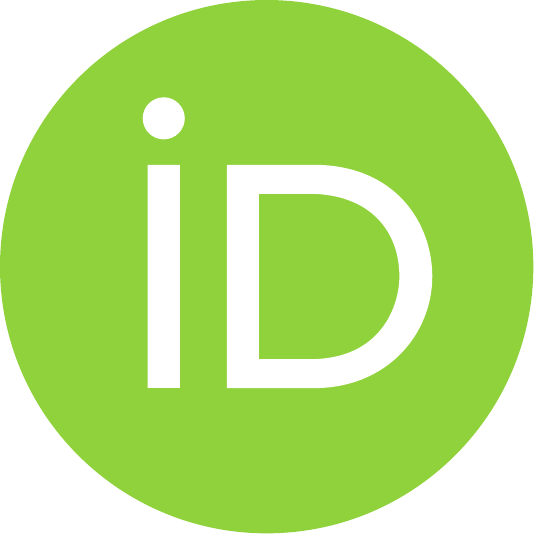}}}

\usepackage[ruled,vlined]{algorithm2e}

\title{Visualizing Evolving Trees}

\author{Kathryn Gray \inst{1}\and Mingwei Li\orcidID{0000-0002-0457-8091} \inst{2}\and Reyan Ahmed\orcidID{0000-0001-6830-9053} \inst{3}\and Stephen Kobourov\orcidID{0000-0002-0477-2724} \inst{1}}

\institute{
Department of Computer Science, University of Arizona\\
\email{ryngray@arizona.edu, kobourov@cs.arizona.edu}\and
Department of Computer Science, Vanderbilt University\\
\email{mingwei.li@vanderbilt.edu}\and
Department of Computer Science, Colgate University\\
\email{rahmed1@colgate.edu}
}

\begin{document}

\maketitle
\pagenumbering{gobble}

\begin{abstract}
Evolving trees arise in many real-life scenarios from computer file systems and dynamic call graphs, to fake news propagation and disease spread. Most layout algorithms for static trees do not work well in an evolving setting (e.g., they are not designed to be stable between time steps). Dynamic graph layout algorithms are better suited to this task, although they often introduce unnecessary edge crossings.  
With this in mind we propose two methods for visualizing evolving trees that guarantee no edge crossings, while optimizing (1) desired edge length realization, (2) layout compactness, and (3) stability. 
We evaluate the two new methods, along with five prior approaches (three static and two dynamic), on real-world datasets using quantitative metrics: stress, desired edge length realization, layout compactness, stability, and running time. The new methods are fully functional and available on github.\footnote{This work was supported in part by NSF grants CCF-1740858, CCF-1712119, and DMS-1839274.}
\end{abstract}

\section{Introduction}\label{section:introduction}

Dynamic graph visualization is used in many fields including  social networks~\cite{gilbert2011communities}, 
bibliometric networks~\cite{van2014visualizing}, 
software engineering~\cite{burch2012visualizing}, and pandemic modeling~\cite{balcan2010modeling};
see the survey by Beck {\it et al.}~\cite{beck2014state}. 
Here we focus on a special case, {\it evolving trees}.
In evolving trees the dynamics are captured only by growth (whereas in general dynamic graphs, nodes and edges can also disappear). While this is a significant restriction of the general dynamic graph model, evolving trees are common in many domains including the
Tree of Life~\cite{tol_web} and the
 Mathematics Genealogy Graph~\cite{math_web}.  An evolving tree can also model disease spread,
where nodes correspond to infected individuals and a new node $v$ is added along with an edge to existing node $u$ if $u$ infected $v$.
Visualizing this process can help us see how the infection spreads, the rate of infection, and to identify ``super-spreaders." 

There are several methods and tools that can be used to visualize evolving trees~\cite{moen1990drawing,cohen1992framework,cohen1995dynamic,archambault2010animation}, however,  most of them have limitations that can impact their usability in this domain. Some represent nodes only as points ignoring labels~\cite{cohen1992framework,cohen1995dynamic}, which makes them less useful in real-life applications where it is important to see what each node represents.
Others utilize the level-by-level approach for drawing hierarchical graphs~\cite{gansner1993technique,north1995incremental}, which does not capture the underlying graph structure well.
Force-directed algorithms tend to better capture the underlying graph structure~\cite{hu2009extending},  
although they may introduce unnecessary edge crossings. 
With this in mind, we propose two methods for drawing crossing-free evolving trees that optimize the following desirable properties:
\begin{enumerate}
\item \textbf{Desired edge length realization:} The Euclidean distance between two nodes $u$ and $v$ in the layout should realize the corresponding pre-specified edge length $l(u,v)$, or be uniform when no additional information is given.
This is important in several domains, e.g., when visualizing phylogenetic trees~\cite{bachmaier2005drawing}, where the edge length represents evolutionary distance between two species.
\item   \textbf{Layout compactness}: The drawing area should be proportional to the total area needed for all the labels. 
A good visualization should have the labeled graph drawn in a compact way~\cite{Nguyen2018}. 
This prevents the trivial solution of scaling the layout until all overlaps and crossings are removed, which can create vast empty spaces in the visualization.

\item \textbf{Stability:}  Between time steps, nodes should move as little as possible. This helps the viewer maintain a mental map of the graph~\cite{misue1995layout}.  If the graph moves around too much, it is difficult to see where new nodes and edges are added and we lose the context of the new node's relation to the rest of the graph. 
\end{enumerate}

\begin{figure}[h]
\centering
    \vspace{-.1cm}\hspace{-.9cm}
    \includegraphics[width=1.06\linewidth]{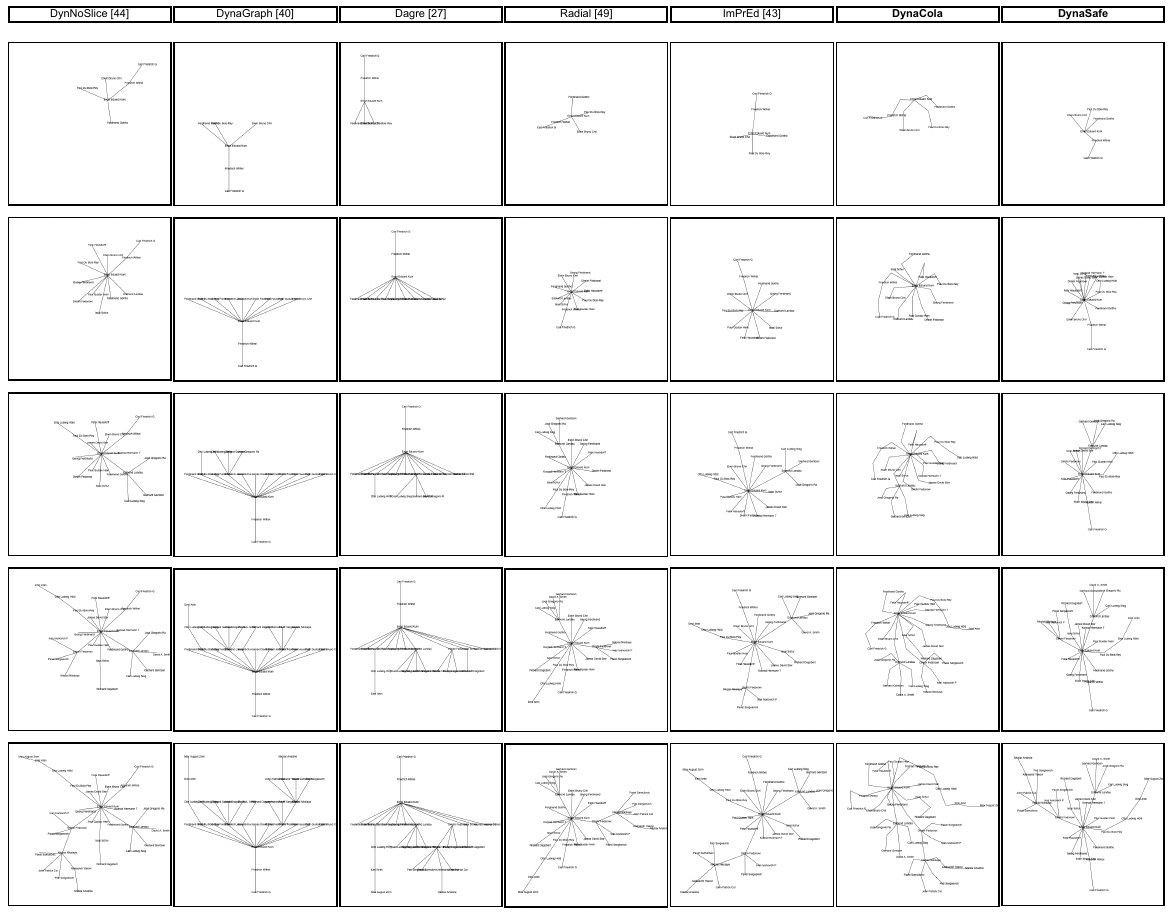}
    \caption{\small Layouts from DynNoSlice, DynaGraph, Dagre, Radial, ImPrEd, DynaCola and DynaSafe of the same evolving math genealogy tree; each row adds six new nodes.}
    \label{fig:dynamic_images_table_labled}
\end{figure}

We propose two force-directed methods that ensure no edge crossings and optimize desired edge length, compactness and stability.
Minimizing edge crossings is important in  graph readability~\cite{Purchase1997}, and since we work with trees, a layout without edge crossings is possible and desirable. 
We use two trees extracted from Tree of Life~\cite{tol_web} and the Mathematics Genealogy~\cite{math_web} projects to demonstrate the new methods and quantitatively evaluate their performance, measuring desired edge length realization, compactness, stability, stress, crossings, and running time. We also evaluate the performance of five earlier methods, showing the two proposed methods perform well overall; see Fig.~\ref{fig:dynamic_images_table_labled}.

\section{Related Work}

Dynamic graph drawing has a long history~\cite{beck2017taxonomy,SkambathT16} and two broad categories: offline and online. In the easier offline setting we assume that all the data about the dynamics is known in advance.
Algorithms for offline dynamic visualization use different approaches including combining all time-slice instances into a single supergraph~\cite{Diehl00,Diehl01,diehl2002graphs}, connecting the same node in consecutive time-slices and optimizing them simultaneously~\cite{erten2003graphael,erten2003simultaneous,forrester2004graphael}, providing animation~\cite{bach2013graphdiaries}, and showing small multiples type visualization~\cite{bach2015small}.
\textit{DynNoSlice} by Simonetto {\it et al.}~\cite{simonetto2018event} is one of the most recent approaches for this setting and is different from the prior methods as it does not rely on discrete time-slices.






Online dynamic graph drawing  deals with the harder problem -- when we do not know in advance what changes will occur. 
One can optimize the current view, given what has happened in the past, but cannot look into the future, as the information is not available.
Cohen {\it et al.}~\cite{cohen1992framework,cohen1995dynamic} and Workman {\it et al.}~\cite{workman2004incremental} describe algorithms for dynamic drawing of trees that place nodes that are equidistant from the root on the same level (same $y$ coordinate). These algorithms do not take edge lengths into consideration, and the hierarchical nature of the layout can lead to exponential differences between the shortest and longest edges.

\begin{figure*}[t]
\centering
\begin{minipage}{.5\textwidth}
  \centering
  \includegraphics[width=.4\linewidth,height=.4\linewidth]{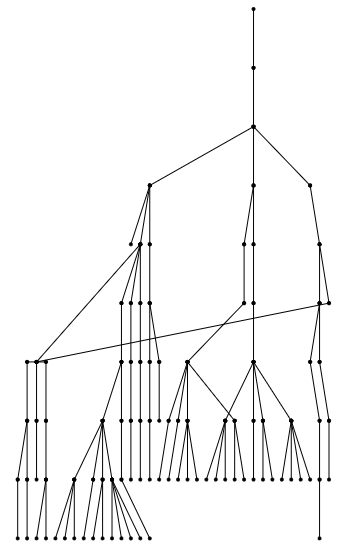}\\(a)
  \label{fig:dynagraph200}
\end{minipage}%
\begin{minipage}{.5\textwidth}
  \centering
  \includegraphics[width=.4\linewidth,height=.4\linewidth]{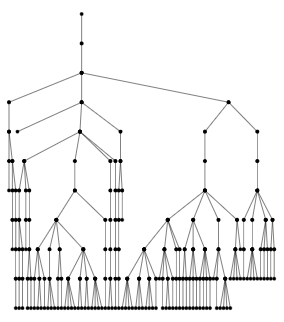}\\(b)
  \label{fig:dagre30}
\end{minipage}
\caption{(a) A DynaDAG layout of a tree with 100 nodes that introduces {\bf edge crossings}. (b) A Dagre layout of a tree with 300 nodes, with {\bf large edge length variability}. Both examples show that {\bf nodes are too close to each other} if labeled. The trees are extracted from the math genealogy dataset.}
\label{fig:hierarchical_weakness}
\end{figure*}

\textit{DynaDAG} is an online graph drawing method for drawing dynamic directed acyclic graphs as hierarchies~\cite{north1995incremental}. 
This method moves nodes between adjacent ranks based on the median sort. It was not specifically developed\footnote{We have used the implementation available in the DynaGraph system: https://www.dynagraph.org/.} for trees and may introduce crossings; see Fig.~\ref{fig:hierarchical_weakness}.
Other approaches for online dynamic graph drawing have maintained the horizontal and vertical position of nodes~\cite{misue1995layout}, used node aging methods~\cite{gorochowski2011using}, and adapted multilevel approaches~\cite{crnovrsanin2015incremental} (using FM$^3$~\cite{hachul2004drawing}). Online approaches have also been implemented on the GPU~\cite{frishman2008online}. However, these methods do not guarantee crossing-free layouts for trees and do not take into account desired edge lengths.


\textit{Dagre} is a multi-phase algorithm for drawing directed graphs based on~\cite{gansner1993technique}. The initial phase finds an optimal rank assignment using the network simplex algorithm. Then it sets the node order within ranks by an iterative heuristic incorporating a weight function and
local transpositions to reduce crossings (via the barycenter heuristic)~\cite{junger20022}. 
However, since Dagre draws graphs in a hierarchical structure, the edge lengths may vary arbitrarily; see Fig.~\ref{fig:hierarchical_weakness}.

\begin{figure}[t]
\centering
\begin{minipage}{.5\textwidth}
  \centering
  \includegraphics[width=.8\linewidth]{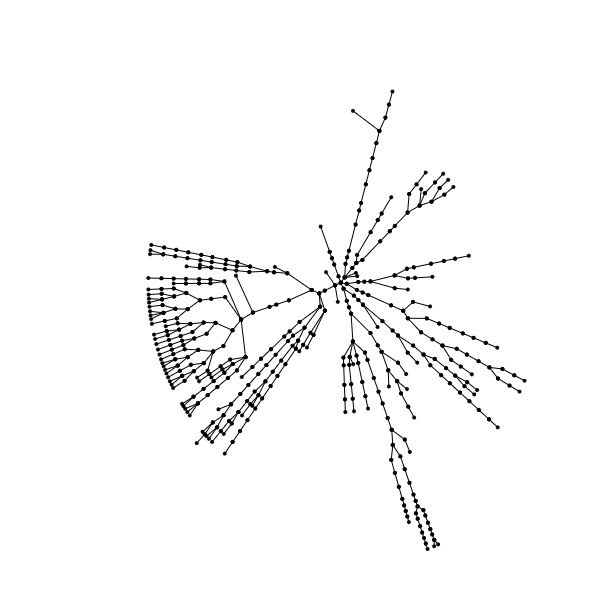}\\(a)
  \label{fig:radial400}
\end{minipage}%
\begin{minipage}{.5\textwidth}
  \centering
  \includegraphics[width=.8\linewidth]{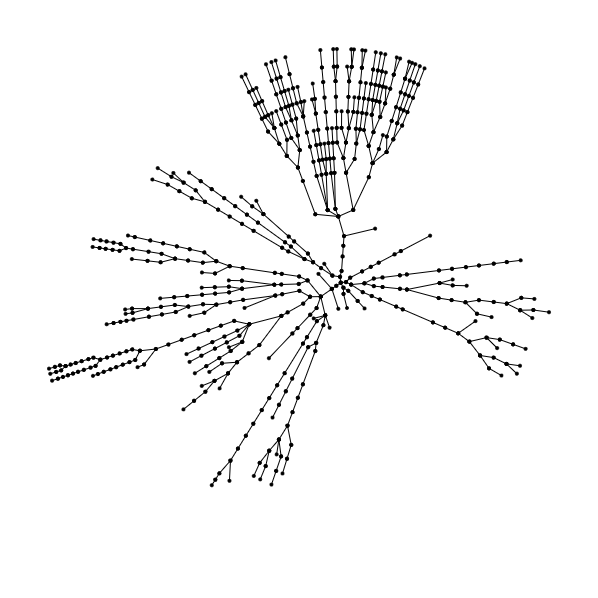}\\(b)
  \label{fig:radial500}
\end{minipage}
\caption{A radial layout of a 400-node evolving tree of life (a) and the layout after adding 100 nodes (b). As most of the growth occurred on the right side, it is easy to see the {\bf instability} -- a lot of movement was needed to accommodate it. Also, {\bf nodes are too close to each other} if labeled.}
\label{fig:unstable_radial}
\end{figure}

The \textit{radial} layout implemented in yFiles~\cite{wiese2004yfiles}  displays each biconnected component in a circular fashion\footnote{We use this radial layout algorithm later for our experiments.}. 
%
Radial layouts were introduced by Kar {\it et al.}~\cite{kar1988heuristic} for static graphs. Dougrusoz {\it et al.}~\cite{dougrusoz1996circular} described an interactive tool for dynamic graph visualization based on the radial layout.
Six and Tollis~\cite{six1999framework} adapted the radial idea to  circular drawings of static biconnected graphs and experimentally showed that their layout has fewer edge crossings.
Kaufmann {\it et al.}~\cite{kaufmann2002maintaining} extended this model to handle dynamic graphs, providing the basis of the yFiles {\it radial} implementation. 
Pavlo {\it et al.}~\cite{pavlo2006parent} adapted the idea to make the radial layout computation parallelizable.
Bachmaier~\cite{bachmaier2007radial} further improved the radial layout algorithm for static graphs by adapting the hierarchical approach~\cite{sugiyama1981methods} to minimize edge crossings.
Radial layout methods have more freedom than the traditional level-by-level tree layout methods. 
Nevertheless, they are still constrained and can result in unstable visualization for dynamic graphs in general and evolving trees in particular; see Fig.~\ref{fig:unstable_radial}.

Force-directed algorithms~\cite{Diehl00,Diehl01,diehl2002graphs,simonetto2018event,simonetto2011impred} 
underlie many static and dynamic graph visualization methods. 
Unlike hierarchical and radial approaches, force-directed algorithms place nodes at arbitrary positions, and so tend to generate more compact layouts that better realize desired edge lengths. By adjusting forces appropriately, one can also generate stable layouts by this approach. 
In particular, ImPrEd, by Simonetto {\it et al.}~\cite{simonetto2011impred}, provides  a force-directed approach to improve a given initial layout, without introducing new edge crossings.
To the best of our knowledge, there are no force-directed methods for evolving trees.




\section{Algorithms for Visualizing Evolving Trees}
Here, we describe two force-directed algorithms for evolving tree visualization, {\it DynaCola} and {\it DynaSafe}, 
that realize desired edge lengths without creating crossings, and optimize compactness and stability. DynaCola avoids edge crossings by creating and maintaining a ``collision region" for each edge. While collision detection/prevention is usually applied to nodes, by carefully applying it to the edges we can prevent all edge crossings.
DynaSafe prevents edge crossings with a ``safe" coordinate update at every step of the algorithm. Before updating a coordinate,  it first checks whether the update  will introduce a crossing and then limits the update magnitude to avoid the crossing.

\subsection{DynaCola}

DynaCola stands for Dynamic Collision, as the algorithm uses the collision forces to prevent edge crossings. This is a force-directed algorithm, augmented with edge-regions used to prevent crossings; see  Algorithm~\ref{alg:DynaCola} in the Appendix. 
Recall that we are gradually growing a tree, one node at a time, while maintaining a crossing-free layout and optimizing desirable properties (desired edge lengths, compactness, stability). 
The DynaCola force-directed algorithm relies on the following forces and is implemented in d3.js~\cite{bostock2011d3}: 

\begin{itemize}
    \item A force $f_E$ for each edge, to realize the desired edge length. The strength of this force is proportional to the difference between the edge distance in the layout and the desired edge length.
    \item A general repulsive force $f_R$ defined for all pairs of nodes and implemented with the Barnes-Hut quad-tree data structure~\cite{barnes1986hierarchical}. This helps realize the global structure of the underlying tree. 
    \item A collision force $f_C$ for each edge, described in details below. This force prevents edge crossings.
    \item A gravitational force $f_G$ that attracts all nodes to the center of mass. This force draws the nodes closer together and improves compactness.
\end{itemize}

To ensure that no edges cross during an update, we define a collision region around each edge: if any edge/node moves too close to another edge, it will be pushed away. 
%
To create a collision region for an edge $e=(u, v)$, we can create collision circles with diameter equal to the length of $e$ for both $u$ and $v$. Then every point of $e$ will be either inside the collision region of $u$ or $v$. However, the sum of all collision regions for all nodes will be unnecessarily large and the layout will not be compact.  
With the help of subdivision nodes along the edges, we can reduce the sum of all collision regions. Let $e=(u, v)$ be an edge in the graph. We use a set of subdivision nodes $V_s = \{v_1, v_2, \cdots, v_k\}$ and replace the edge $(u, v)$ by a set of edges $E_s = \{(u, v_1), (v_1, v_2), \cdots, (v_k, v)\}$. We assign the desired edge length of an edge in $E_s$ equal to $l(u,v)/|E_s|$, where $l(u,v)$ is the desired edge length.   
In general, the number of subdivision nodes per edge should be a small constant $n_s$ (by default $n_s=1$), since the complexity of the algorithm increases as $n_s$ increases. 
Also, note that $n_s$ determines the number of bends per edge (no bends when $n_s=0$, one bend when $n_s=1$, and so on). Note that, the collision force does not follow any hard constraint, even after having a collision region edge crossings may happen. If existing edges introduce crossings, then we roll back to previous crossing-free coordinates.
When a new node is added to the tree, a new edge also is added, with one of its endpoints already placed.  To place the new node, we randomly sample a set of 100 nearby points at a distance equal to the desired edge length, trying to find a crossing-free position. If we cannot find such a suitable point, we gradually reduce the distance and repeat the search until we find a crossing free position. 
Once the new node has been placed, we subdivide its adjacent edge as described above. 

\subsection{DynaSafe}
DynaSafe stands for Dynamic Safety, as the algorithm prioritizes safe moves and will not make a move if it introduces an edge crossing, see  Algorithm~\ref{alg:DynaSafe} in Appendix.  DynaSafe is also a force-directed algorithm, however, it differs from DynaCola as it draws straight-line edges (rather than edges with bends).  
The algorithm utilizes the following forces and is implemented in  d3.js~\cite{bostock2011d3} 

\begin{itemize}
\item A force $f_E$ for each edge that is similar to DynaCola.
\item A stress-minimizing force $f_S$ on every pair of nodes not connected by an edge, used to improve global structure. The desired distance is the shortest-path distance between the pair, 
and the magnitude of the force is proportional to the difference between the realized and desired distance. 
\item A repulsive force that is similar to DynaCola.
\item A gravitational force that is similar to DynaCola.
\end{itemize}


DynaSafe prevents edge crossings from occuring at any time by updating the coordinates safely: if the proposed new coordinate of a node introduces crossings, we gradually reduce the magnitude of the movement until the crossing is avoided.  
To place the new node, we randomly sample a set of 100 nearby 
points to find a crossing-free position for its adjacent edge. If we cannot find a crossing free position using the sample points, we continuously reduce the edge length until we find a crossing free position.
Once the node is added, an iteration of force-directed algorithm optimizes the layout (again without introducing crossings).

By the nature of force-directed algorithms, after one phase of force computations each node has a proposed new position. 
Before moving any node to its proposed new position, we check that the move is ``safe," i.e., it does not introduce a crossing.
If the movement of a node introduces any crossings, then the magnitude of the move is set to $p$\% of the original movement. This is repeated (if needed) at most $q$ times, and if the crossing is still unavoidable then the node is not moved in this phase. By default  $p=0.8$ and $q=12$.

\section{Experimental Evaluation}\label{section:evaluation}
We evaluate DynaCola and DynaSafe, along with five earlier methods: DynNoSlice, DynaGraph, Dagre, Radial, and ImPrEd. We use two evolving trees to visually compare the results, as well quantitatively evaluate the desired properties.

\subsection{Datasets}

We use two real-world datasets to extract evolving trees for our experiments.

\medskip\noindent\textbf{The Tree of Life:}  captures the evolutionary progression of life on Earth~\cite{tol_web}. The underlying data is a tree structure with a natural time component. As a new species evolves, a new node in the tree is added. 
The edges give the parent-child relation of the nodes, where the parent is the original species, and the child is the new species. We use a subset of this graph with 500 nodes. The maximum node degree of this tree is 5, and the radius is 24.

\medskip\noindent\textbf{The Mathematics Genealogy:} shows advisor-advisee relationships in the world of mathematics, stretching back to the middle ages~\cite{math_web}. 
The dataset includes the thesis titles, students, advisors, dates, and number of descendants. The total number of nodes is  around 260,000 and is continuously updated. While this data is not quite a tree (or even connected, or planar), we extract a subset to create a tree with 500 nodes. The maximum node degree of this tree is 5 and the radius is 14.


\subsection{Evaluation Metrics}

We use standard metrics for each of our desired properties: desired edge length preservation, compactness, and stability. Additionally, we compute the stress of the drawing and the number of crossings. This gives a total of five quantitative measures. For each of these measures we define a loss function as follows:

\medskip\noindent\textbf{Desired Edge Length (DEL):}  To measure how close the realized edge lengths are to the desired edge lengths, we find the mean squared error between these two values. 
 Given the  desired edge lengths $\{l_{ij}: (i,j) \in E\}$ and coordinates of the nodes $X$ in the computed layout, we evaluate with the following  formula:
 
    \begin{align}
    \text{\texttt{Desired edge length loss}} &= \sqrt{ \frac{1}{|E|} \sum\limits_{(i,j) \in E}\;  
    \left(\frac{||X_i - X_j|| - l_{ij}}{l_{ij}}\right)^2} \label{eq:loss-desired-edge-length}
    \end{align}
    
    This measures the root mean square of the relative error as in~\cite{ahmed2020gd}, producing a non-negative number, with $0$ corresponding to a perfect realization. For DynaCola we subdivide the edges, to compute DEL, we set the length of the subdivided edges such that the summation of the length of the subdivided edges is equal to the length of the original edge.

 \medskip\noindent\textbf{Compactness:} To measure the compactness of each layout, we use the ratio between the drawing area and the sum of the areas for all labels~\cite{SemanticWordClouds}. We assume that a label is at most 16 characters, as we abbreviate longer labels. 
The sum of the areas for all labels gives the minimum possible area needed to draw all labels without overlaps (ignoring any space needed for edges).  The area of the actual drawing is given by the smallest bounding rectangle, once the drawing has been scaled up until there are no overlapping labels.
Once we have this scaled drawing, we find the positions of the nodes with the largest and smallest x and y values ($X_{max,0}$, $X_{min, 0}, X_{max,1}$ and $X_{min,1}$).  Using these values we calculate the area of the bounding rectangle.


\begin{align}
    \text{\texttt{Compactness loss}} &= \frac{(X_{max,0}-X_{min,0})(X_{max,1}-X_{min,1})}{ \sum\limits_{v \in V} \text{label\_area}(v) }
     \label{eq:loss-Compactness}
    \end{align}
    
This formula produces a non-negative number; the ideal value for this measure is $1$ and  corresponds to a perfect space utilization.

 \medskip\noindent\textbf{Stability:} To measure stability, we consider how much each of the nodes moved after adding a new node. We then sum the movements of all nodes over all time steps. Since different algorithms use different amounts of drawing areas, we divide the value by the drawing area to normalize the results. This measure is similar to that used in DynNoSlice~\cite{simonetto2018event}, but since DynNoSlice does not use time slices, it is closer to the measure found in~\cite{brandes2011quantitative}:
 
 \begin{align}
    \text{\texttt{Stability loss}} &= \frac{\sum_{v \in V} \sum_{t=1}^{T-1} ||X_v(t+1) - X_v(t)||}{(X_{max,0}-X_{min,0})(X_{max,1}-X_{min,1})}
     \label{eq:loss-Stability}
    \end{align}
    
 Here, $T$ is the maximum time (500 in our two datasets). This formula produces a non-negative number; the ideal value is $0$ and corresponds to a perfectly stable layout (no movement of any already placed nodes).
 %

 \medskip\noindent\textbf{Stress: } This measure evaluates the global quality of the layout, looking at the differences between the realized distance between any pair of nodes and the actual distance between them.  This measure is used in a variety of graph drawing algorithms~\cite{simonetto2018event,brandes2011quantitative,gansner2004graph}: 
 
 \begin{align}
     \text{\texttt{Stress loss}} &= \frac{\bigg(\sum_{i\ne j} \big(D_{i,j}-||X_i -X_j||\big)^2\bigg)^{1/2}}{\sum_{i\ne j} ||X_i -X_j||}
     \label{eq:loss-Stress}
 \end{align}
 
 Here, $D_{i,j}$ is the shortest path distance in the graph.
 This formula produces a non-negative number; the ideal value is $0$ and corresponds to a perfect embedding (that captures all graph distances by the realized Euclidean distances).

\medskip\noindent\textbf{Edge Crossings:} 
Finally, we measure the number of edge crossings in each of the outputs.  Note that our algorithms DynaSafe and DynaCola enforce  ``no edge crossings" as a hard constraint.  However,  DynNoSlice and DynaGraph do not have such a constraint and so can and indeed do, introduce crossings.  Therefore we include the number of edge crossings for a complete comparison.

\subsection{Experimental Setup}

We compare these algorithms to five previous algorithms: DynNoSlice, DynaGraph, Dagre,  Radial, and ImPrEd.  We note that while Dagre, Radial, and ImPrEd are not specifically designed for dynamic graphs, they can be modified for this purpose. Specifically, we can use the layout of a tree at step $i$ to initialize the layout of the tree at step $i+1$, add the new edge, and update the layout.

We consider the simplest case for  the desired edge length by using a uniform length of 100 for all edges.  This is a necessary parameter for our algorithms DynaCola and DynaSafe, but only needed in the other four algorithms in order to compute the desired edge length measure. To be able to compare our methods to the other four (that do not take desired edge length into account), we set the desired edge length equal to the average edge length obtained in the layout. We then normalize these values for a fair comparison.








The performance of DynNoSlice depends heavily on two parameters, $\tau$ and $\delta$ that must be tuned. With the help of the authors, we found $\tau = 16$ and $\delta = 4$ worked well for our 500-node trees. 
The performance of ImPrEd depends on two parameters: repulsion force and the number of iterations. The default values of repulsion force and the number of iterations are equal to one and 200 respectively. We have used the default values. The larger the number of iterations is, the better the output of ImPrEd is. However, the running time increases as the number of iterations increases. We keep the number of iterations equal to 200 since it already takes more than 4 hours to compute the 500-node trees.
The performance of DynaCola depends on the number of subdivision nodes $n_s$; we use $n_s=1$ for the experiments.
For the radial layout algorithm, we have used the default settings in the  yFiles~\cite{wiese2004yfiles} implementation. The other algorithms are also used with their default settings. 
We have implemented our algorithms in d3.js~\cite{bostock2011d3}. For other algorithms, we have used the default API. All experiments are conducted in a machine that has macOS 11.3.1 operating system, a 2.3 GHz 8-core Intel core i9 processor, and 32 GB 2667 MHz DDR4 memory.


\subsection{Results}
\begin{figure}[t]
\centering
    \hspace{-.9cm}\includegraphics[width=1.06\linewidth]{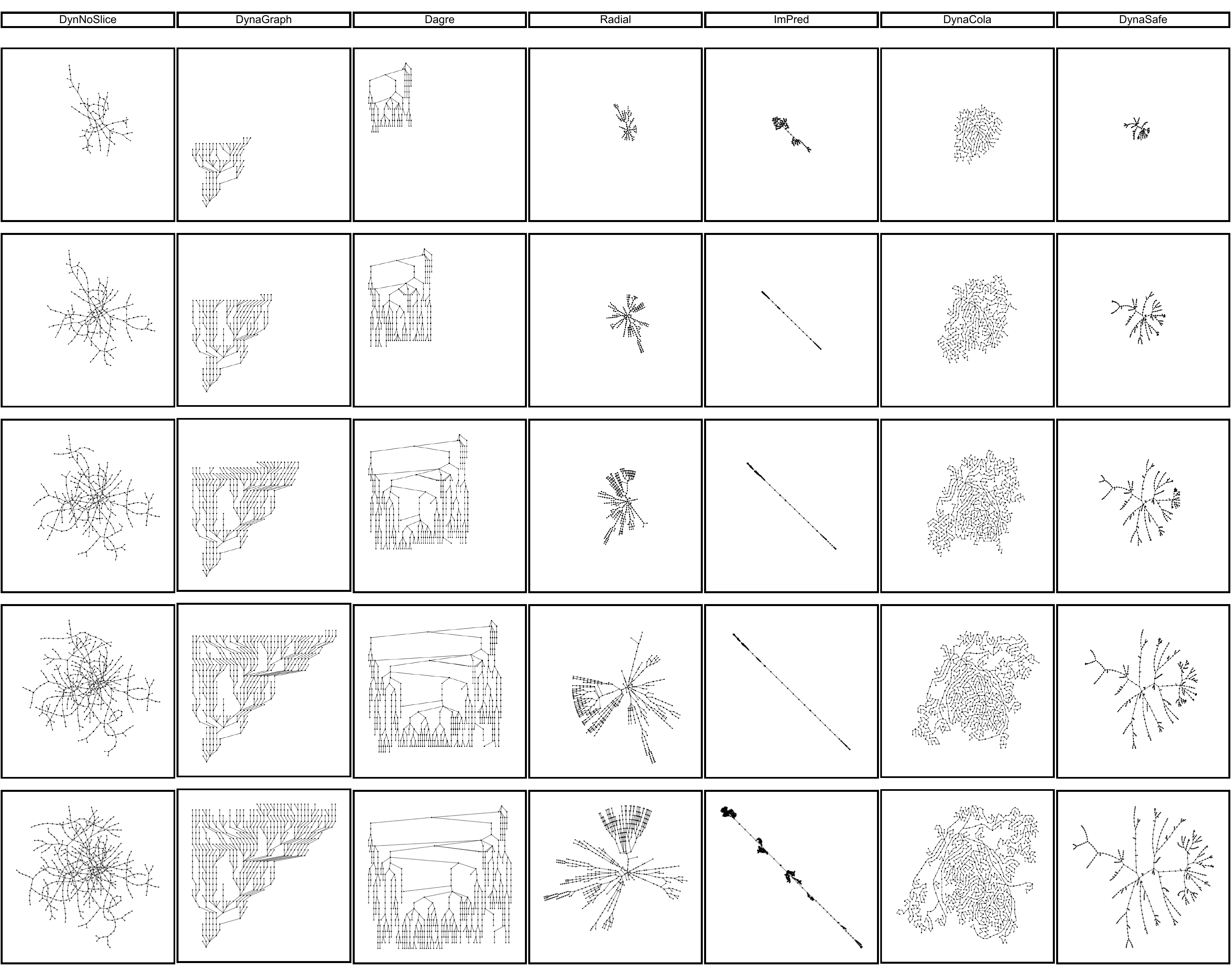}
    \caption{Layouts obtained by the seven methods for the tree of life dataset. 
    }
    \label{fig:dynamic_images_table_tol}
\end{figure}

Both the visual and quantitative results indicate that the two new methods perform well overall; see Fig.~\ref{fig:dynamic_images_table_tol}.

\medskip\noindent\textbf{Desired Edge Lengths: } The quantitative results are shown in Table~\ref{fig:del_math}. We use green to show the  best results and yellow  for the second best and indicate that DynNoSlice, DynaCola, and ImPrEd perform well. For the math genealogy 500-node tree, ImPrEd is the best. However, both DynNoSlice and ImPrEd have significantly larger running times (measured in hours, rather than minutes or seconds) as discussed below. Moreover, while ImPrEd does well on the math genealogy graph, it does not do well on the tree of life graph.  For the math genealogy 500-node tree, DynaCola is the second best and DynNoSlice is third. For the tree of life dataset, DynaCola is the best, DynNoSlice is the second best and DynaSafe is third. DynaGraph has the worst performance -- not surprising given that it is a hierarchical layout, which is forced to use some very long edges near the root.

\begin{table}[h]
    \begin{center}
    \begin{tabular}{|c|c|c|c|c|c|c|c|}
    \hline
     Nodes & DynNoSlice & DynaGraph & Dagre & Radial & ImPrEd & DynaCola & DynaSafe \\ 
     \hline \hline
     100 MG & 0.37691 & 1.95933 & 0.682568 & 0.653853 & \cellcolor{green!50}0.219103 & \cellcolor{yellow!50}0.282521 & 0.589865 \\  
     \hline
     200 MG & 0.36552 & 1.95179 & 0.679827 & 0.640628 & \cellcolor{green!50}0.213615 & \cellcolor{yellow!50}0.270322 & 0.575430 \\
     \hline
     300 MG & 0.35007 & 1.94213 & 0.666440 & 0.63058 & \cellcolor{green!50}0.204821 & \cellcolor{yellow!50}0.253877 & 0.564747 \\
     \hline
     400 MG & 0.34402 & 1.93822 & 0.646479 & 0.619203 & \cellcolor{green!50}0.193037 & \cellcolor{yellow!50}0.243184 & 0.553141 \\
     \hline
     500 MG & 0.33377 & 1.91979 & 0.639766 & 0.592694 & \cellcolor{green!50}0.182071 & \cellcolor{yellow!50}0.237139 & 0.548756 \\
           \hline\hline
     100 TOL & \cellcolor{yellow!50}0.21675 & 1.28710 & 0.448483 & 0.448205 & 0.45402 & \cellcolor{green!50}0.158071 & 0.411747 \\  
     \hline
     200 TOL& \cellcolor{yellow!50}0.21972 & 1.37271 & 0.494261 & 0.460161 & 0.49120 & \cellcolor{green!50}0.166190 & 0.443935 \\
     \hline
     300 TOL& \cellcolor{yellow!50}0.23986 & 1.40404 & 0.510473 & 0.481034 & 0.52016 & \cellcolor{green!50}0.176748 & 0.453332 \\
     \hline
     400 TOL& \cellcolor{yellow!50}0.25597 & 1.45660 & 0.553543 & 0.510352 & 0.59326 & \cellcolor{green!50}0.183856 & 0.470334 \\
     \hline
     500 TOL& \cellcolor{yellow!50}0.26652 & 1.52650 & 0.581648 & 0.530249 & 0.61093 & \cellcolor{green!50}0.189373 & 0.485759 \\
     \hline
    \end{tabular}
    \end{center}
    \hspace{-.5cm}\caption{Desired edge lengths of math genealogy tree (MG) and tree of life (TOL).}
    \label{fig:del_math}
\end{table}

\medskip\noindent\textbf{Compactness: } 
The quantitative results are shown in Table~\ref{fig:compactness_math} and indicate that DynNoSlice outperforms the rest of the algorithms. DynaCola is second best and DynaSafe is third. Here, Dagre has the worst performance. Although DynNoSlice performs well, it introduces many edge crossings as discussed later. DynaCola layouts have higher compactness than DynaSafe. The absence of stress-related force allows placing nodes closer even if the graph theoretic distance is higher. Consider a path, the layout will be a straight line if stress is minimized. However, a zig-zag layout will provide better compactness.

\begin{table}[h]
    \begin{center}
    \begin{tabular}{|c|c|c|c|c|c|c|c|}
    \hline
     Nodes & DynNoSlice & DynaGraph & Dagre & Radial & ImPrEd & DynaCola & DynaSafe \\ 
     \hline \hline
     100 MG & \cellcolor{green!50}85.60 & 192.07 & 219.20 & 153.53 & 161.23 & \cellcolor{yellow!50}124.90 & 147.80 \\  
     \hline
     200 MG & \cellcolor{green!50}87.94 & 196.29 & 224.54 & 162.80 & 161.20 & \cellcolor{yellow!50}130.87 & 153.31 \\
     \hline
     300 MG & \cellcolor{green!50}95.24 & 201.41 & 225.86 & 169.34 & 171.92 & \cellcolor{yellow!50}137.00 & 159.87 \\
     \hline
     400 MG & \cellcolor{green!50}98.89 & 206.48 & 227.74 & 175.07 & 181.20 & \cellcolor{yellow!50}145.47 & 169.68 \\
     \hline
     500 MG & \cellcolor{green!50}106.82 & 208.39 & 236.94 & 192.53 & 187.94 & \cellcolor{yellow!50}149.43 & 174.93 \\
     \hline\hline
     100 TOL & \cellcolor{green!50}96.46 & 196.79 & 223.43 & 179.05 & 179.20 & \cellcolor{yellow!50}147.82 & 160.58 \\  
     \hline
     200 TOL & \cellcolor{green!50}100.24 & 214.29 & 231.38 & 183.06 & 218.29 & \cellcolor{yellow!50}154.10 & 169.38 \\
     \hline
     300 TOL & \cellcolor{green!50}110.56 & 216.16 & 239.98 & 190.82 & 329.27 & \cellcolor{yellow!50}157.19 & 170.04 \\
     \hline
     400 TOL & \cellcolor{green!50}119.98 & 233.85 & 255.76 & 203.92 & 416.27 & \cellcolor{yellow!50}167.52 & 194.28 \\
     \hline
     500 TOL & \cellcolor{green!50}126.85 & 235.72 & 272.82 & 214.09 & 528.01 & \cellcolor{yellow!50}173.94 & 196.03 \\
     \hline
    \end{tabular}
    \end{center}
    \caption{Compactness of math genealogy tree (MG) and the tree of life (TOL).}
    \label{fig:compactness_math}
\end{table}

\medskip\noindent\textbf{Stability: } The quantitative results are shown in Table~\ref{fig:instability_math} and idicate that DynaCola does best. DynaGraph is second, and DynaSafe is third.
The radial layout performs worse in this metric because it rotates the subtrees as more edges are added.

\begin{table}[h]
    \begin{center}
    \begin{tabular}{|c|c|c|c|c|c|c|c|}
    \hline
     Nodes & DynNoSlice & DynaGraph & Dagre & Radial & ImPrEd & DynaCola & DynaSafe \\ 
     \hline \hline
     100 MG & 0.001584 & \cellcolor{green!50}0.001393 & 0.0016502 & 0.001998 & 0.001530 & \cellcolor{green!50}0.001348 & 0.001459 \\  
     \hline
     200 MG & 0.000752 & 0.000447 & 0.0012497 & 0.001839 & 0.000598 & \cellcolor{green!50}0.000264 & \cellcolor{yellow!50}0.000410 \\
     \hline
     300 MG & 0.000577 & \cellcolor{yellow!50}0.000227 & 0.0010083 & 0.001450 & 0.000437 & \cellcolor{green!50}0.000225 & 0.000295 \\
     \hline
     400 MG & 0.000249 & \cellcolor{yellow!50}0.000190 & 0.0009504 & 0.001203 & 0.000391 & \cellcolor{green!50}0.000164 & 0.000216 \\
     \hline
     500 MG & 0.000037 & \cellcolor{yellow!50}0.000014 & 0.0007591 & 0.001047 & 0.000026 & \cellcolor{green!50}0.000011 & 0.000019 \\
     \hline 
     \hline
     100 TOL & 0.000437 & \cellcolor{yellow!50}0.000139 & 0.001241 & 0.003609 & 0.000491 & \cellcolor{green!50}0.000105 & 0.000163 \\  
     \hline
     200 TOL & 0.000323 & \cellcolor{yellow!50}0.000125 & 0.001196 & 0.003408 & 0.008305 & \cellcolor{green!50}0.000097 & 0.000146 \\
     \hline
     300 TOL & 0.000263 & \cellcolor{yellow!50}0.000099 & 0.001163 & 0.003174 & 0.005305 & \cellcolor{green!50}0.000072 & 0.000106 \\
     \hline
     400 TOL & 0.000235 & \cellcolor{yellow!50}0.000073 & 0.001136 & 0.002490 & 0.001937 & \cellcolor{green!50}0.000064 & 0.000101 \\
     \hline
     500 TOL & 0.000199 & \cellcolor{yellow!50}0.000071 & 0.000834 & 0.001941 & 0.000810 & \cellcolor{green!50}0.000052 & 0.000101 \\
     \hline
    \end{tabular}
    \end{center}
    \caption{Stability of math genealogy tree (MG) and tree of life (TOL).}
    \label{fig:instability_math}
\end{table}

\medskip\noindent\textbf{Stress: } The quantitative results are shown in  Table~\ref{fig:stress_math}. In general, DynaSafe does much better on this measure  than the rest. Again, ImPrEd performs well for the regular-shaped math genealogy tree but does not perform well for the tree of life. For the tree of life DynNoSlice is second and DynaCola third. 
DynaGraph and Dagre perform the worst in this metric due to the limitations inherent in the hierarchical layout.
Note that the stress is normalized, so the numbers are comparable.

\begin{table}[h]
    \begin{center}
    \begin{tabular}{|c|c|c|c|c|c|c|c|}
    \hline
     Nodes & DynNoSlice & DynaGraph & Dagre & Radial & ImPrEd & DynaCola & DynaSafe \\ 
     \hline \hline
     100 MG & 113.10 & 150.59 & 230.75 & 125.37 & 89.43 & \cellcolor{yellow!50}76.51 & \cellcolor{green!50}49.44 \\  
     \hline
     200 MG & 161.45 & 186.17 & 250.98 & 172.05 & \cellcolor{yellow!50}120.45 & 151.74 & \cellcolor{green!50}68.42 \\
     \hline
     300 MG & 179.48 & 264.90 & 286.77 & 205.98 & \cellcolor{yellow!50}148.02 & 184.47 & \cellcolor{green!50}94.05 \\
     \hline
     400 MG & 186.68 & 292.66 & 286.83 & 262.09 & \cellcolor{yellow!50}173.92 & 227.98 & \cellcolor{green!50}107.56 \\
     \hline
     500 MG & 249.61 & 393.11 & 396.72 & 314.93 & \cellcolor{yellow!50}203.54 & 291.39 & \cellcolor{green!50}109.00 \\
     \hline     \hline
     100 TOL & 136.52 & 210.06 & 263.98 & 192.64 & 163.02 & \cellcolor{yellow!50}128.45 & \cellcolor{green!50}59.77 \\  
     \hline
     200 TOL & 165.75 & 262.24 & 325.65 & 243.59 & 349.28 & \cellcolor{yellow!50}201.76 & \cellcolor{green!50}65.10 \\
     \hline
     300 TOL & 181.62 & 305.15 & 369.57 & 287.93 & 427.09 & \cellcolor{yellow!50}220.63 & \cellcolor{green!50}81.81 \\
     \hline
     400 TOL & 254.20 & 328.59 & 398.11 & 317.28 & 509.32 & \cellcolor{yellow!50}306.89 & \cellcolor{green!50}93.99 \\
     \hline
     500 TOL & 285.19 & 400.81 & 461.43 & 374.02 & 593.19 & \cellcolor{yellow!50}351.23 & \cellcolor{green!50}119.77 \\
     \hline
    \end{tabular}
    \end{center}
    \caption{Stress scores of math genealogy tree (MG) and tree of life (TOL).}
    \label{fig:stress_math}
\end{table}

\medskip\noindent\textbf{Edge Crossings: } The quantitative results are shown in Table~\ref{fig:crossings_math}. There are five winners here -- the five algorithms that prevent any edge crossings: DynaCola, DynaSafe, Radial, Dagre, and ImPrEd.
DynNoSlice and DynaGraph do introduce some crossings.

\begin{table}[h]
    \begin{center}
    \begin{tabular}{|c|c|c|c|c|c|c|c|}
    \hline
     Nodes & DynNoSlice & DynaGraph & Dagre & Radial & ImPrEd & DynaCola & DynaSafe \\ 
     \hline \hline 
     100 MG & 43 & \cellcolor{yellow!50}8 & \cellcolor{green!50}0 & \cellcolor{green!50}0 & \cellcolor{green!50}0 & \cellcolor{green!50}0 & \cellcolor{green!50}0 \\  
     \hline
     200 MG & 82 & \cellcolor{yellow!50}11 & \cellcolor{green!50}0 & \cellcolor{green!50}0 & \cellcolor{green!50}0 & \cellcolor{green!50}0 & \cellcolor{green!50}0 \\
     \hline
     300 MG & 168 & \cellcolor{yellow!50}11 & \cellcolor{green!50}0 & \cellcolor{green!50}0 & \cellcolor{green!50}0 & \cellcolor{green!50}0 & \cellcolor{green!50}0 \\
     \hline
     400 MG & 217 & \cellcolor{yellow!50}13 & \cellcolor{green!50}0 & \cellcolor{green!50}0 & \cellcolor{green!50}0 & \cellcolor{green!50}0 & \cellcolor{green!50}0 \\
     \hline
     500 MG & 277 & \cellcolor{yellow!50}13 & \cellcolor{green!50}0 & \cellcolor{green!50}0 & \cellcolor{green!50}0 & \cellcolor{green!50}0 & \cellcolor{green!50}0 \\
     \hline\hline
     100 TOL & \cellcolor{yellow!50}21 & \cellcolor{green!50}0 & \cellcolor{green!50}0 & \cellcolor{green!50}0 & \cellcolor{green!50}0 & \cellcolor{green!50}0 & \cellcolor{green!50}0 \\  
     \hline
     200 TOL & \cellcolor{yellow!50}67 & \cellcolor{green!50}0 & \cellcolor{green!50}0 & \cellcolor{green!50}0 & \cellcolor{green!50}0 & \cellcolor{green!50}0 & \cellcolor{green!50}0 \\
     \hline
     300 TOL & \cellcolor{yellow!50}106 & \cellcolor{green!50}0 & \cellcolor{green!50}0 & \cellcolor{green!50}0 & \cellcolor{green!50}0 & \cellcolor{green!50}0 & \cellcolor{green!50}0 \\
     \hline
     400 TOL & \cellcolor{yellow!50}176 & \cellcolor{green!50}0 & \cellcolor{green!50}0 & \cellcolor{green!50}0 & \cellcolor{green!50}0 & \cellcolor{green!50}0 & \cellcolor{green!50}0 \\
     \hline
     500 TOL & \cellcolor{yellow!50}231 & \cellcolor{green!50}0 & \cellcolor{green!50}0 & \cellcolor{green!50}0 & \cellcolor{green!50}0 & \cellcolor{green!50}0 & \cellcolor{green!50}0 \\
     \hline
    \end{tabular}
    \end{center}
    \caption{Edge crossings of math genealogy tree (MG) and tree of life (TOL).}
    \label{fig:crossings_math}
\end{table}


\medskip\noindent\textbf{Running time: } The Radial layout has the lowest running time, taking 34.93 seconds and 28.03 seconds, respectively, to draw the 500-node math genealogy tree and tree of life. On the other end, DynNoSlice is the slowest algorithm, taking more than 6 hours to draw the 500-node trees.
Both DynNoSlice and ImPrEd take significantly longer running time compared to other algorithms. ImPrEd takes more than four hours to draw the 500-node trees.
Our two new methods are not as fast as the Radial algorithm and not as slow as DynNoSlice, taking about 5 minutes on the 500-node trees.
Due to space limitations, we provide more details of running time in the Appendix.

\section{Discussion and Limitations}

While there are many algorithms and tools for drawing static trees, only a few can handle dynamic trees well. 
Among those, even fewer takes edge labels into account while also preventing edge crossings.
With this in mind, we described two methods that give better, readable layouts for  evolving trees. 
We compared these two algorithms with others that have been set up for dynamic trees. 
With respect to the criteria that we have put forward, our algorithms match or exceed each of these algorithms.
Fully functional prototypes and videos showing them in action are available online 
{\small \url{https://ryngray.github.io/dynamic-trees/}}. 
Source code and all experimental data can be found on github 
{\small \url{https://github.com/abureyanahmed/evolving_tree}}.

Naturally, our work comes with several limitations that could be addressed in future work.

\textbf{Anticipating the Future}: 
Currently, the two new methods, DynaCola and DynaSafe perform well for evolving trees, where growth is the only type of change. 
A natural question is whether these algorithms can be generalized to the more challenging problems of online dynamic tree visualization.
Answering such question may need more precise modeling of the graph dynamics.
Even though online dynamic graph drawing assumes no knowledge about the actual changes to the graph in the future, some prior knowledge of the graph may be available or predictable in advance. 
For example, knowing the expected depth or size of a tree or maximum degree of nodes (e.g., from domain knowledge about the specific type of graph) may help the layout algorithm  reserve enough space for growth.
In general, we anticipate that if one can model the evolving dynamics of the graph (e.g., probabilistically), incorporating knowledge of such dynamics into the drawing algorithm may help improve the resultant drawing; conversely, carefully defining compatible graph dynamics for a particular drawing algorithm will also allow us to identify the limitations of the given algorithm.

\textbf{Multi-level Label  Display}:
For simplicity, in this work we assume labels to be always shown in the drawing in a fixed font size.
In practice, however, labels may come with different levels of importance and different desired font size.
In that case, one might prefer to see only important labels displayed first in a zoomed out view of the graph, and later see more labels when zooming in.
Incorporate such multi-level label display into the node placement strategy seems like an interesting and relevant problem.

\textbf{Finding Desired Properties}: 
We have proposed two different algorithms to solve the same evolving tree visualization problem, and each is associated with different benefits.
Finding a continuous \textit{spectrum} of algorithms with tunable parameters to balance the multiple desired properties would provide  more flexibility.
On the other hand, a careful human-subjects study may also help  prioritize existing properties of the drawing, or help  identify new desired properties from the specific tasks.

\textbf{Considering More Dynamic Datasets}:
The datasets we considered are evolving in nature. For example, in the math genealogy dataset, once an advisee gets related to an advisor, the relationship remains forever. Although we considered only evolving trees, our ideas can be applied to datasets where the elements may get deleted. Applying the algorithms on more dynamic datasets remains future work.

\subsection*{Acknowledgements}  We thank the authors of DynNoSlice and DynaGraph
for their assistance with running and tuning algorithms. We also thank yFiles whose radial layout implementation we use in the evaluation.


\bibliography{references}
\bibliographystyle{splncs04}

\appendix
\section*{\LARGE Appendix}

We provide some details about the forces that DynaCola and DynaSafe use in each iteration. DynaCola uses a collision force $f_C$.
The goal of this collision force is to remove edge crossings and label overlaps.
To define the collision force, each node is assigned a collision circle centered around the node. When a node $u$ enters the collision circle of another node $v$ then a repulsive force $f_r$ acts between $u$ and $v$:
$f_r = K/d \cdot I_{d \leq r}(d)$, 
where $I_{d \leq r}$ is an indicator function for the condition $d \leq r$,  i.e $I_{d \leq r}(d) = 1$ if $d \leq r$ and $0$ otherwise.
Here $K$ is a constant, by default equal to the area of the initial drawing.
The force $f_r$ is activated when the distance $d$ from $u$ to $v$ is less than or equal to the collision circle radius $r$. The value of $f_r$ is reciprocal to $d$. We apply this collision force to every node.

Both DynaCola and DynaSafe use an edge force $f_E$; the edge force works to maintain the desired edge lengths. 
In the initial embedding, the edge lengths are equal to the desired lengths. If we only apply the collision forces we might remove all overlaps at the expense of drastically modifying the edge lengths. To avoid this, we combine edge forces with collision forces. 
For every edge we apply either a repulsive force $f^e_r = K/d \cdot I_{d < l_e}(d)$ (when the edge is compressed) or an attractive force $f^e_a = K d \cdot I_{d > l_e}(d)$ (when the edge is stretched), determined by the indicator function.
The force is proportional/reciprocal to distance $d$.

The algorithms use a general repulsive force and a gravitational force as well. These forces have different directions, but they are similar in principle to the collision force and the edge force.

\begin{algorithm}[]
\SetAlgoLined
\KwResult{A crossing free layout, maintaining the desired properties}
 Let $u, v = e$\;
 Let $u \in T$\;
 Find a crossing free position for $v$ by random sampling\;
 Reduce the edge length of $e$ if necessary\;
 Add $n_s$ subdivision nodes for $e$\;
 Add the new edges and dummy nodes to $T$\;
 \For{each node $w$ of $T$}{
  $\Delta w = 0$\;
 }
 \For{$i \in \{1, 2, \cdots, nIters\}$}
 {
  Compute the collision circle radius\;
  \For{each edge $e'$ of $T$}{
   Let $u', v' = e'$\;
   $\Delta u', \Delta v' += f_E(e')$\;
  }
  \For{each node $w$ of $T$}{
   $\Delta w += F_R(w)$\;
   $\Delta w += F_C(w)$\;
   $\Delta w += F_G(w)$\;
  }
  \For{each node $w$ of $T$}{
   $X_w += \Delta w$\;
  }
 }
 \caption{DynaCola(Current tree $T$, Current layout $\Gamma_T$, New edge $e$, Number of iterations $nIters$, Number of subdivision nodes $n_s$)}
 \label{alg:DynaCola}
\end{algorithm}

\begin{algorithm}[h]
\SetAlgoLined
\KwResult{A crossing free layout, maintaining the desired properties}
 Let $(u, v) = e$\;
 Let $u \in T$\;
 Find a crossing free position for $v$ by random sampling\;
 Reduce the edge length of $e$ if necessary\;
 Add the new edge to $T$\;
 \For{each node $w$ of $T$}{
  $\Delta w = 0$\;
 }
 \For{$i \in \{1, 2, \cdots, nIters\}$}
 {
  \For{each edge $e'$ of $T$}{
   Let $(u', v') = e'$\;
   $\Delta u', \Delta v' += f_E(e')$\;
  }
  \For{each node $w$ of $T$}{
   $\Delta w += F_R(w)$\;
   $\Delta w += F_G(w)$\;
  }
  \For{each node pair $w, x$}{
   $\Delta w, \Delta x += F_S(w, x)$\;
  }
   \For{each node $w$ of $T$}{
   Reduce $\Delta w$ by $p$\% for at most $q$ times\;
   Update $X_w$ by $\Delta w$ if no crossing occurs\;
  }
 }
 \caption{DynaSafe(Current tree $T$, Current layout $\Gamma_T$, New edge $e$,  Number of iterations $nIters$)}
 \label{alg:DynaSafe}
\end{algorithm}

\begin{figure}[h]
    \centering
    \includegraphics[width=1.0\linewidth]{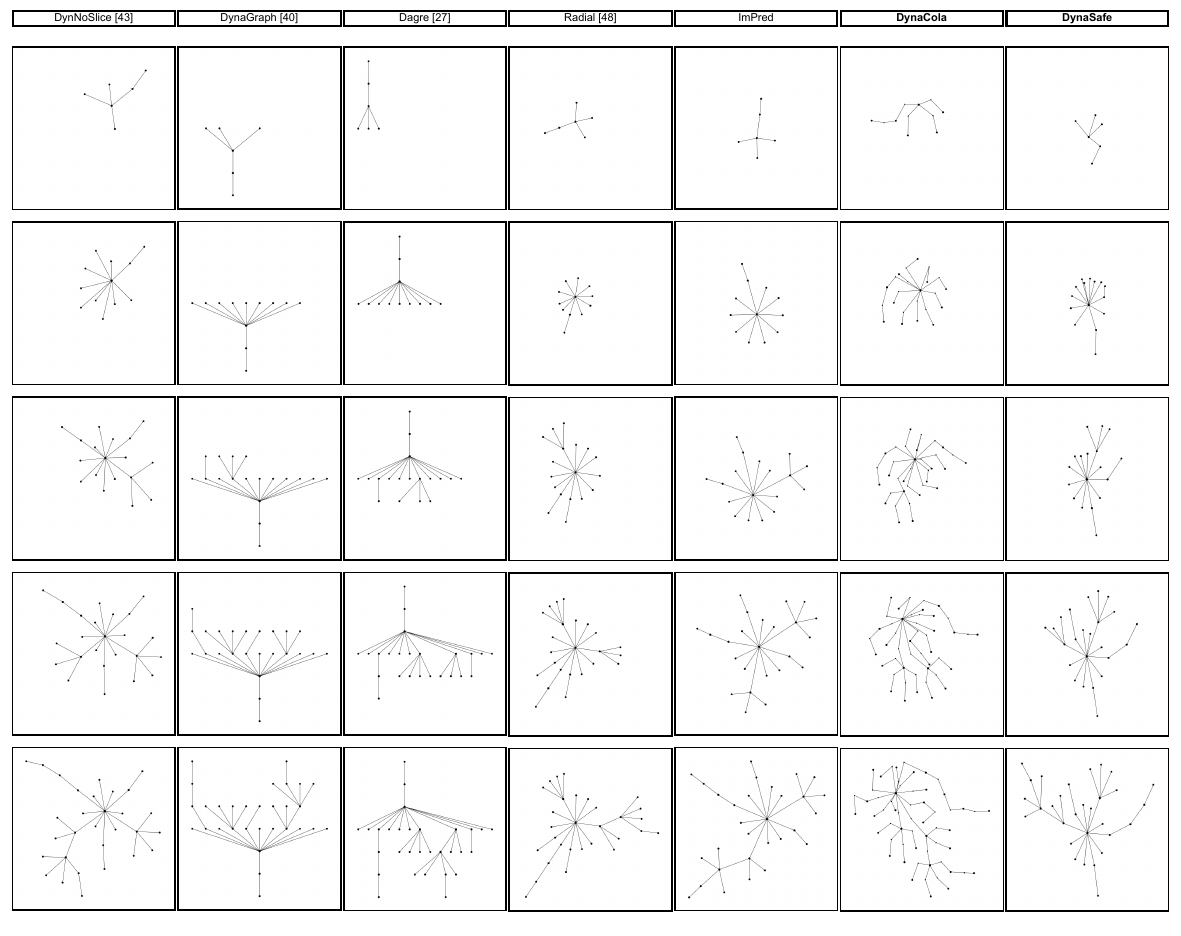}
    \caption{The unlabeled layouts obtained by DynNoSlice, DynaGraph, Dagre, Radial, ImPrEd, DynaCola, and DynaSafe of the same evolving math genealogy tree; each row adds six new nodes.}
    \label{fig:dynamic_images_table}
\end{figure}

Fig.~\ref{fig:dynamic_images_table_labled} shows a small evolving tree with labels, as obtained by the seven algorithms: DynNoSlice, DynaGraph, Dagre, Radial, ImPrEd, DynaCola, and DynaSafe. Fig.~\ref{fig:dynamic_images_table} shows the same layouts without the labels, allowing us to see the structure a bit better.

\begin{figure}[h]
    \hspace{-.5cm}\includegraphics[width=1.1\linewidth]{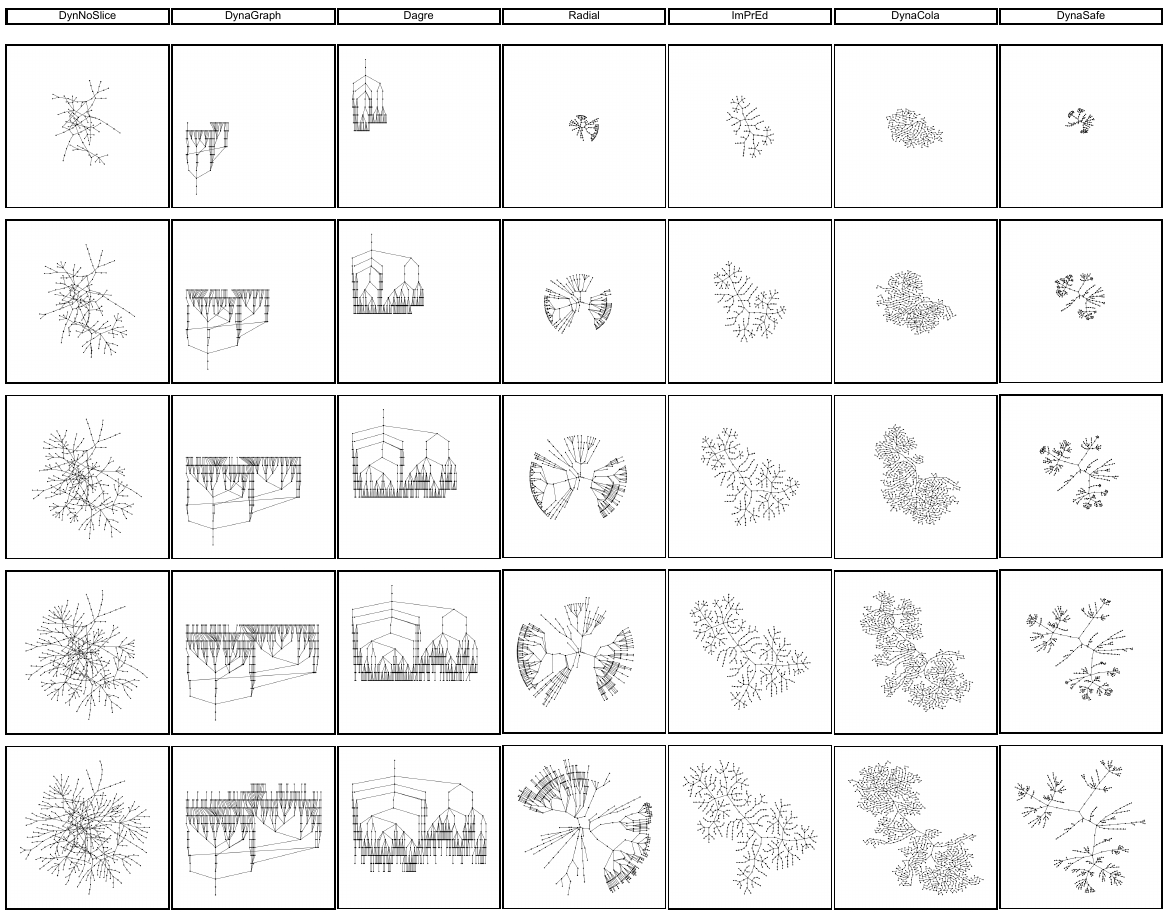}
    \caption{The layouts of different algorithms of the same evolving math genealogy trees at different steps.}
    \label{fig:dynamic_images_table_math}
\end{figure}

Fig.~\ref{fig:dynamic_images_table_tol} shows the layouts of the tree of life, obtained by DynNoSlice, DynaGraph, Dagre, Radial, ImPrEd, DynaCola and DynaSafe. Fig.~\ref{fig:dynamic_images_table_math} shows the layouts of math genealogy tree, obtained by DynNoSlice, DynaGraph, Dagre, Radial, ImPrEd, DynaCola and DynaSafe. 

Tab.~\ref{fig:run_time} shows the running times of DynNoSlice, DynaGraph, Dagre, Radial, ImPrEd, DynaCola and DynaSafe.

\begin{table}[h]
    \begin{center}
    \begin{tabular}{|c|c|c|c|c|c|c|c|}
    \hline
     Graph & DynNoSlice & DynaGraph & Dagre & Radial & ImPred & DynaCola & DynaSafe \\ 
     \hline
     100 MG & 536.86 & \cellcolor{yellow!50}23.01 & 31.39 & \cellcolor{green!50}4.01 & 206.65 & 53.27 & 103.61 \\  
     \hline
     200 MG & 1253.38 & \cellcolor{yellow!50}36.02 & 51.04 & \cellcolor{green!50}11.42 & 1151.15 & 176.92 & 149.28 \\  
     \hline
     300 MG & 3376.19 & \cellcolor{yellow!50}43.19 & 78.15 & \cellcolor{green!50}17.39 & 5451.75 & 201.82 & 267.71 \\  
     \hline
     400 MG & 9382.27 & \cellcolor{yellow!50}57.23 & 96.14 & \cellcolor{green!50}27.20 & 9460.37 & 286.27 & 326.16 \\  
     \hline
     500 MG & 24804.39 & \cellcolor{yellow!50}72.77 & 117.87 & \cellcolor{green!50}34.93 & 15497.94 & 368.48 & 398.15 \\  
     \hline
     \hline
     100 TOL & 647.33 & \cellcolor{yellow!50}9.12 & 13.78 & \cellcolor{green!50}3.71 & 191.52 & 27.81 & 31.39 \\
     \hline
     200 TOL & 1297.64 & \cellcolor{yellow!50}13.04 & 19.26 & \cellcolor{green!50}9.52 & 1120.00 & 43.01 & 59.74 \\
     \hline
     300 TOL & 3529.79 & \cellcolor{yellow!50}19.37 & 28.03 & \cellcolor{green!50}15.38 & 4281.03 & 67.13 & 87.10 \\
     \hline
     400 TOL & 9104.15 & \cellcolor{yellow!50}25.28 & 41.03 & \cellcolor{green!50}21.91 & 11937.18 & 79.41 & 119.47 \\
     \hline
     500 TOL & 22212.30 & \cellcolor{yellow!50}36.28 & 52.49 & \cellcolor{green!50}28.03 & 22206.56 & 99.30 & 153.89 \\
     \hline
    \end{tabular}
    \end{center}
    \caption{Running time of different algorithms in seconds.}
    \label{fig:run_time}
\end{table}

DynaCola performs slightly better compared to DynaSafe. However, DynaCola layouts introduce bends. DynaCola layouts may be suitable where compact layouts are preferred. Since Dynasafe has better stress and does not create bends, the layouts are suitable for scenarios where the Euclidean distance between the endpoints of an edge provides valuable information. For example, visualizing phylogenetic trees~\cite{bachmaier2005drawing}, where the edge length represents the evolutionary distance between two species.

\end{document}